\documentclass[apj]{emulateapj}

\usepackage{epsfig}
\usepackage{lscape,graphicx,rotating}
\usepackage{apjfonts}

\begin{document}

\title{GRBs\,070429B and 070714B: The High End of the Short-Duration 
Gamma-Ray Burst Redshift Distribution\altaffilmark{1}}

\author{S.~Bradley Cenko\altaffilmark{2}, Edo Berger\altaffilmark{3,4,5},
        Ehud Nakar\altaffilmark{6}, Mansi M.~Kasliwal\altaffilmark{7}, 
        Antonio Cucchiara\altaffilmark{8}, Shri R.~Kulkarni\altaffilmark{7},
        Eran Ofek\altaffilmark{7}, Derek B.~Fox\altaffilmark{8}, 
        Fiona A.~Harrison\altaffilmark{2}, Arne Rau\altaffilmark{7},
        Paul A.~Price\altaffilmark{9}, Avishay Gal-Yam\altaffilmark{10},
        Michael A.~Dopita\altaffilmark{11}, Bryan E.~Penprase\altaffilmark{12}}
\altaffiltext{1}{Some of the data presented herein were obtained at the 
        W.~M.~Keck Observatory, which is operated as a scientific partnership 
        among the California Institute of Technology, the University of 
        California and the National Aeronautics and Space Administration. The 
        Observatory was made possible by the generous financial support of the 
        W.~M.~Keck Foundation.}
\altaffiltext{2}{Space Radiation Laboratory, MS 220-47,
        California Institute of Technology, Pasadena, CA 91125}
\altaffiltext{3}{Observatories of the Carnegie Institute of Washington,
        813 Santa Barbara Street, Pasadena, CA 91101}
\altaffiltext{4}{Princeton University Observatory, Peyton Hall, Ivy Lane,
        Princeton, NJ 08544}
\altaffiltext{5}{Hubble Fellow}
\altaffiltext{6}{Theoretical Astrophysics, California Institute of Technology, 
        Pasadena, CA 91125}
\altaffiltext{7}{Division of Physics, Mathematics, and Astronomy, 105-24,
        California Institute of Technology, Pasadena, CA 91125}
\altaffiltext{8}{Department of Astronomy and Astrophysics, Pennsylvania
        State University, 525 Davey Laboratory, University Park, PA 16802}
\altaffiltext{9}{Institute for Astronomy, 2680 Woodlawn Drive, Honolulu, HI, 
        96822}
\altaffiltext{10}{Benoziyo Center for Astrophysics, Weizmann Institute of 
        Science, 76302 Rehovot, Israel}
\altaffiltext{11}{Research School of Astronomy and Astrophysics, ANU, Cotter 
        Road, Weston Creek, ACT, 2611, Australia}
\altaffiltext{12}{Department of Physics and Astronomy, Pomona College, 610 
        North College Avenue, Claremont, CA 91711}

\email{cenko@srl.caltech.edu}

\slugcomment{\textit{ApJL} submitted}

\shorttitle{GRBs\,070429B and 070714B}
\shortauthors{Cenko \textit{et al.}}


\newcommand{\Swift}{\textit{Swift}}
\newcommand{\KW}{\textit{Konus-Wind}}
\newcommand{\HST}{\textit{Hubble Space Telescope}}
\newcommand{\gmos}{\textit{GMOS}}
\newcommand{\mgii}{\ion{Mg}{2} $\lambda \lambda$ 2796, 2803}
\newcommand{\mgi}{\ion{Mg}{1} $\lambda$ 2852}
\newcommand{\oii}{[\ion{O}{2} $\lambda$ 3727]}

\begin{abstract}
We present optical spectra of the host galaxies of the short-duration 
gamma-ray burst GRB\,070429B and the likely short-duration with extended
emission GRB\,070714B.  In both cases, we find a single emission line that we 
identify as [\ion{O}{2}] $\lambda$ 3727 at $z \sim 0.9$.  Both events are more 
distant than any previous short-duration GRB with a secure host association 
from the sub-arcsecond position of an optical afterglow.  GRBs\,070429B and 
070714B provide strong evidence in support of our previous claims in Berger et 
al.~that a significant fraction of short-duration hosts ($\gtrsim 33\%$) reside 
at $z > 0.7$.  We discuss the implications of the existence this population on 
the energetics of short-duration GRBs, as well as on progenitor models.  In 
the context of the degenerate binary merger scenario, such events 
require progenitor systems with a range of lifetimes and disfavor progenitor 
models with a long, narrow lifetime distribution.
\end{abstract}


\section{Introduction}
\label{sec:intro}
Much like their long-duration cousins, the discovery of the first X-ray 
\citep{gso+05,bpp+06}, optical \citep{hwf+05}, and radio \citep{bpc+05} 
afterglows of short-duration ($\Delta t \lesssim 2$\,s) gamma-ray bursts (GRBs) 
has driven great progress in recent years.  Based primarily on the discovery of 
the first four afterglows, GRB\,050509B ($z = 0.225$; \citealt{gso+05,bpp+06}),
GRB\,050709 ($z = 0.1606$; \citealt{vlr+05,hwf+05,ffp+05}), GRB\,050724 
($z = 0.257$; \citealt{bcb+05,bpc+05,gbp+06}), and GRB\,051221 ($z = 0.5465$;
\citealt{sbk+06,bgc+06}), a general picture of short-duration GRB hosts quickly 
emerged.  Unlike long-duration events, short-duration GRBs occur in
both early- and late-type galaxies, all at relatively low redshift (compared
to $\bar{z} \approx 3$ for long-duration \Swift\ GRBs; \citealt{bkf+05,jlf+06}).
If, as is widely believed, short-duration GRBs result from coalescence of a 
compact binary (neutron star -- neutron star [NS-NS] or neutron star -- black
hole [NS-BH]; \citealt{elp+89}), these nearby events imply a long progenitor
delay time ($\tau > 4$\,Gyr; \citealt{ngf06}), inconsistent with the properties
of such systems observed in the Milky Way.

In a recent work, \citet{bfp+07} have searched the afterglow error circles of 
nine additional short-duration bursts without an established redshift, 
identifying a population of faint ($R \sim 23$ -- 26\,mag) putative host 
galaxies.  Spectroscopy of the brightest four of these galaxies indicate they 
lie at $0.4 < z < 1.1$, while a comparison with field galaxies indicates the 
fainter hosts likely reside at $z \geq 1$.  The implications of the existence 
of a high-redshift ($z > 0.7$) population of short-duration bursts would be 
profound, in terms of the implied isotropic energy release 
($E_{\gamma,\mathrm{iso}} \sim 10^{50}$ -- $10^{51}$\,erg), the progenitor delay
times (both shorter and a broader distribution), and the possibility of
detection with ground-based gravitational wave detectors.

In this \textit{Letter}, we present spectra of the host galaxies
of two recent GRBs: the short-duration GRB\,070429B and the likely
short-duration with extended emission GRB\,070714B.  Both events have detected 
optical afterglows, providing both a secure host association and a secure 
redshift, thereby overcoming the primary limitation of our previous statistical 
study.  Both hosts lie at $z \sim 0.9$, providing strong evidence in favor of a 
substantial population of high-redshift short-duration bursts.  We conclude by 
discussing the impact of this population on short-duration burst progenitor 
models.  

\section{GRB\,070429B}
\label{sec:070429}
The Burst Alert Telescope (BAT; \citealt{bbc+05}) on board \Swift\ 
detected and localized GRB\,070429B at 03:09:04 UT \citep{GCN.6358}.  The 
duration of the prompt emission was measured to be $t_{90} = 0.5 \pm 0.1$\,s, 
placing the event firmly in the short-duration category \citep{GCN.6365}.  
Fitting a power-law with index $\Gamma = 1.71 \pm 0.23$, \citet{GCN.6365} 
measure a prompt fluence of $F_{\gamma} = (6.3 \pm 1.0) \times 10^{-8}$\,erg 
cm$^{-2}$ in the 15 -- 150\,keV band.

\begin{figure}[t]
        \centerline{\plotone{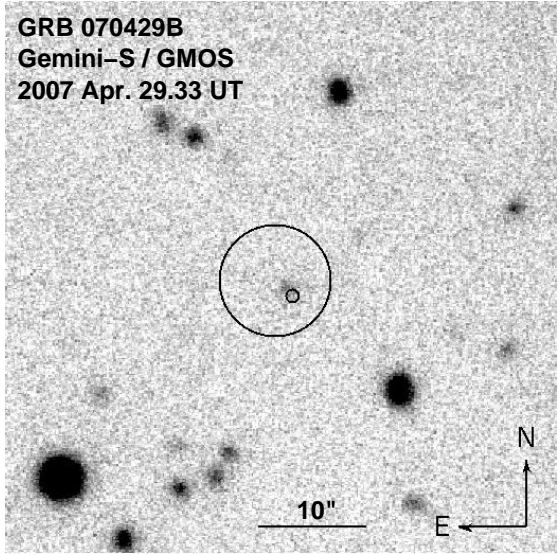}}
        \caption[Optical imaging of the field of GRB\,070429B]
        {Optical imaging of the field of GRB\,070429B.  The X-ray error circle
        is shown as the large black circle \citep{GCN.6360}, while the
        location of the optical afterglow is indicated with the small black
        circle \citep{GCN.7145}.  The size of the small circle is indicative of
        the astrometric error in the afterglow location.  The likely host
        galaxy is offset from the afterglow location by less than $0.6\arcsec$.}
\label{fig:0429_finder}
\end{figure}

The fading X-ray afterglow was identified by the X-ray Telescope (XRT;
\citealt{bhn+05}) on board \Swift\ in ground analysis at location $\alpha = 
21^{\mathrm{h}} 52^{\mathrm{m}} 03\farcs82$, $\delta = -38^{\circ} 49\arcmin
42\farcs2$ (J2000.0) with a 90\%-confidence uncertainty of $5\farcs1$ 
\citep{GCN.6360}.  We imaged the position of the X-ray afterglow with the 
Gemini Multi-Object Spectrograph (GMOS; \citealt{hab+03}) mounted on the 
8-m Gemini South telescope beginning at UTC 07:59:55 29 April ($\sim 4.8$\,h 
after the trigger).  We identified a single faint, extended object inside the 
XRT error circle at coordinates $\alpha = 21^{\mathrm{h}} 52^{\mathrm{m}} 
03\farcs729$, $\delta = -38^{\circ} 49\arcmin 42\farcs84$ (J2000.0), with an 
uncertainty of $0\farcs3$ (Fig.~\ref{fig:0429_finder}; \citealt{GCN.6370}).  
An additional, fainter source inside the XRT error circle was identified by 
\citet{GCN.6372} at coordinates $\alpha = 21^{\mathrm{h}} 52^{\mathrm{m}} 
03\farcs4$, $\delta = -38^{\circ} 49\arcmin 43\farcs1$ (J2000.0).  The field 
was observed by several other facilities in the following months, yet no 
variability was detected in either source (e.g.~\citealt{GCN.6604,GCN.7140}). 

Subsequent re-analysis of the initial observations from the \Swift\ UV-Optical 
telescope (UVOT; \citealt{rkm+05}) beginning 230\,s after the burst trigger 
revealed a marginal source in the $U$, $B$, and $V$ filters at $\alpha = 
21^{\mathrm{h}} 52^{\mathrm{m}} 03\farcs68$, $\delta = -38^{\circ} 49\arcmin
43\farcs6$ (J2000.0) which subsequently faded away \citep{GCN.7145}.  While 
none of the individual UVOT detections was particularly significant, summing 
all seven UVOT filters from 591 to 2661\,s after the trigger results in a 
3.9-$\sigma$ detection \citep{GCN.7145}.  We therefore accept that this source 
is the optical afterglow of GRB\,070429B.

The UVOT afterglow falls only $0\farcs6$ from the centroid of the candidate 
host first proposed by \citet{GCN.6370} (Fig.~\ref{fig:0429_finder}).  
Because the angular extent of the galaxy in our Gemini image is about
1\arcsec, the UVOT afterglow position is directly coincident with this
galaxy.  Using the cumulative galaxy counts in the HUDF and GOODS surveys
\citep{bsk+06}, we estimate the probability of an $R \lesssim 23$\,mag galaxy 
falling within $0\farcs6$ of the afterglow location to be $\sim 3.1 \times 
10^{-3}$.  On the other hand, the chance alignment of the other object 
identified in the XRT error circle by \citet{GCN.6372} is as high as 0.3.  
It is therefore likely that this source is the host galaxy of GRB\,070429B.  

We obtained a spectrum of this object with the Low Resolution Imaging 
Spectrograph (LRIS; \citealt{occ+95}) mounted on the 10-m Keck I telescope on 
the night of 12 August 2007.  Two 1800 s exposures were taken at high air mass
($\sim 1.9$) with the 0\farcs7 slit.  We employed a dichroic at 5600\,\AA\ to 
split the beam in two.  The 400/8500 grating on the red side provides coverage 
from 5600 -- 8500 \AA, with a dispersion of 1.86\,\AA\ pixel$^{-1}$.  The 
400/3400 grism was responsible for the dispersion on the blue end, resulting 
in coverage from the atmospheric cut-off up to $\sim$ 5600\,\AA\ and a 
dispersion of 1.09 \AA\ pixel$^{-1}$.  

Spectra were reduced in the IRAF\footnote{IRAF is distributed by the
National Optical Astronomy Observatory, which is operated by the
Association for Research in Astronomy, Inc., under cooperative agreement with
the National Science Foundation.} environment using standard
routines.  Dithered spectra were subtracted to remove residual
sky lines.  Cosmic rays were removed using the LA Cosmic routine
\citep{v01}.  The resulting spectrum was extracted optimally \citep{h86} and 
wavelength calibration was performed first relative to calibration lamps and 
then tweaked based on night sky lines.  Both air-to-vacuum and heliocentric 
corrections were then applied to all spectra.  Extracted spectra were divided 
through by a smoothed flux standard to remove narrow band ($< 50$\,\AA) 
instrumental effects \citep{b99}.  Flux calibration was performed relative to 
the spectrophotometric standard Feige 11 \citep{s96}.  To account for slit
losses, we scaled the resulting spectrum to match the measured photometric 
$g^{\prime}$- and $R$-band magnitudes \citep{GCN.7140}.  Finally, we de-reddened 
the spectrum to account for the modest Galactic extinction: $E(B-V) = 0.027$
\citep{sfd98}.

\begin{figure}[t]
        \epsscale{1.2}
        \centerline{\plotone{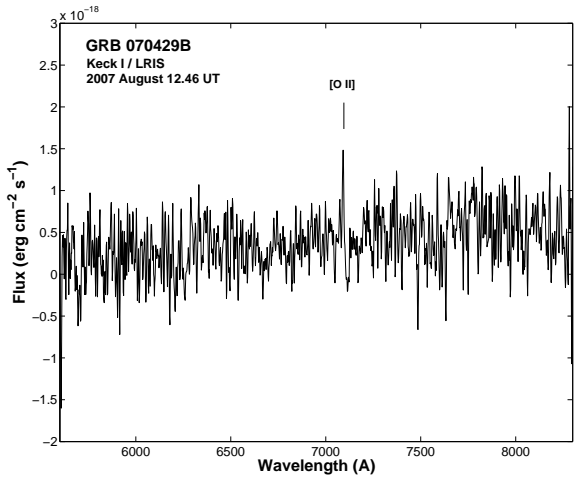}}
        \caption{Keck I/LRIS spectrum of the likely host galaxy of GRB\,070429B,
        smoothed with a 3 pixel boxcar.  We detect a single emission line that
        we identify as [\ion{O}{2}] $\lambda$ 3727 at $z = 0.9023 \pm 0.0003$.}
\label{fig:0429_spec}
\end{figure}

The resulting spectrum of the host galaxy of GRB\,070429B is shown in 
Figure~\ref{fig:0429_spec}.  We find a single faint emission line at 
$\lambda = 7091.2 \pm 0.6$\,\AA.  Due to the lack of blueward features, we 
identify this line as [\ion{O}{2}] $\lambda$ 3727 at a redshift of $z = 
0.9023 \pm 0.0002$.  Fitting the line with a Gaussian profile, we measure a 
flux of $F = (3.8 \pm 1.2) \times 10^{-17}$\,erg cm$^{-2}$ s$^{-1}$.  Adopting 
the most recent $\Lambda$CDM parameters for a flat universe from WMAP ($H_{0} = 
73$\,km s$^{-1}$ Mpc$^{-1}$; $\Omega_{\mathrm{m}} = 0.24$, $\Omega_{\Lambda} = 1 - 
\Omega_{\mathrm{m}}$; \citealt{sbd+07}), and using the transformation from 
\citet{k98}, we infer a star formation rate of $1.1 \pm 0.5$\,M$_{\odot}$ 
yr$^{-1}$.  Correcting for the rest-frame $B$-band luminosity, we find 
a specific star formation rate of $\phi = 2.5 \pm 1.1$\,M$_{\odot}$ yr$^{-1}$ 
L$_{*}^{-1}$.  We caution, however, these values are only lower limits, as they 
do not incorporate any extinction native to the host.  These 
results are in overall agreement with the findings in the 
GCN\footnote{Gamma-Ray Burst Circulars Network: 
http://gcn.gsfc.nasa.gov/gcn3\_archive.html} circular of \citet{GCN.7140}.

\section{GRB\,070714B}
\label{sec:070714}

GRB\,070714B was detected by the \Swift-BAT at 04:59:29 UT on 17 July 2007
\citep{GCN.6620}.  Using the best-fit power-law spectrum with index $\Gamma = 
1.36 \pm 0.19$, the 15 -- 150\,keV fluence in the prompt emission was measured 
to be $F_{\gamma} = (5.1 \pm 0.3) \times 10^{-7}$\,erg cm$^{-2}$.  While the 
duration of the prompt emission is significantly longer than the canonical 
long-short divide ($t_{90} = 64 \pm 5$\,s), the burst light curve is comprised 
of several short spikes (duration $\sim 3$\,s) with a long, soft tail 
\citep{GCN.6623}.  The spectral lag is consistent with zero \citep{GCN.6631}, 
a property typically observed in short-duration bursts \citep{nb06}.  Like GRBs 
050709 and 050724, GRB\,070714B therefore appears likely to be a member of the 
short-duration bursts with extended emission (e.g.~\citealt{bcb+05}).
However, together with the uncertainty regarding the duration, we caution that 
a small number of long-duration (i.e.~massive star progenitor) \Swift\ events 
have been observed with zero lag \citep{gnb+06}.

A fading optical afterglow was discovered inside the XRT error circle
\citep{GCN.6627} by several groups \citep{GCN.6621,GCN.6630,GCN.6635}.  The
latest flux measurements and most precise astrometric position
were reported by \citet{GCN.6652}, who find an $R$-band magnitude of 25.5\,mag 
and location $\alpha = 03^{\mathrm{h}} 51^{\mathrm{m}} 22\farcs3$,
$\delta = +28^{\circ} 17\arcmin 50\farcs8$ ($\pm 0\farcs4$; J2000.0).
The probability of a chance association for such a precise localization 
($\approx 0\farcs4$) is quite small, $\sim 3 \times 10^{-3}$ \citep{bsk+06}.
We obtained two 2100\,s dithered spectra of the host galaxy of GRB\,070714B
with LRIS on the night of 13 August 2007.  These spectra were reduced in an
identical manner as those obtained for GRB\,070429B (\S \ref{sec:070429}).  
Flux calibration was performed relative to the spectrophotometric standard
star BD+17$^{\circ}$4708 \citep{og83,bg04}.

\begin{figure}[t]
        \epsscale{1.2}
        \centerline{\plotone{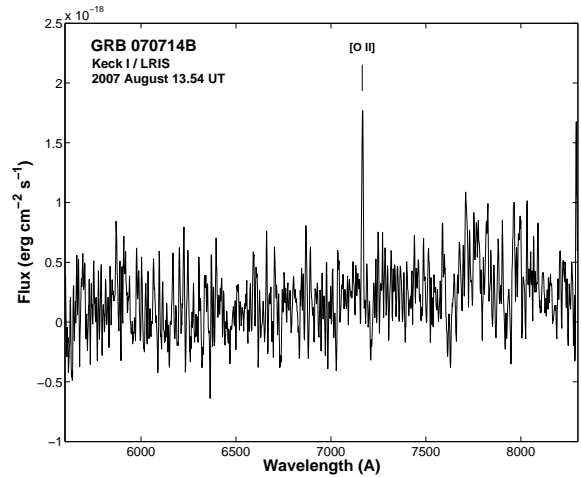}}
        \caption{Keck I/LRIS spectrum of the likely host galaxy of GRB\,070714B,
        smoothed with a 3 pixel boxcar.  We detect a single emission line that
        we identify as [\ion{O}{2}] $\lambda$ 3727 at $z = 0.9224 \pm 0.0001$.}
\label{fig:0714_spec}
\end{figure}

The resulting spectrum is shown in Figure~\ref{fig:0714_spec}.  We find a single
bright emission line at $\lambda = 7166.2 \pm 0.4$, which we again identify as
[\ion{O}{2}] $\lambda$ 3727 at $z = 0.9225 \pm 0.0001$.  This confirms
the spectroscopic redshift first proposed by \citet{GCN.6836} in the GCN
circulars.  We measure a flux of $1.6 \pm 0.3 \times 10^{-17}$\,erg cm$^{-2}$ 
s$^{-1}$ for this line, corresponding to a lower limit on the star formation 
rate of $0.5 \pm 0.2$\,M$_{\odot}$ yr$^{-1}$ \citep{k98}.  The corresponding
specific star formation rate for GRB\,070714B is $\phi=8.3 \pm 2.8$\,M$_{\odot}$ 
yr$^{-1}$ L$_{*}^{-1}$.

\section{Discussion}
\label{sec:discussion}

In summary, we present the redshifts of the host galaxies of the short-duration 
GRB\,070429B ($z = 0.902$) and the short-duration with extended emission 
GRB\,070714B ($z = 0.922$).  These are the two highest redshifts for 
short-duration GRBs with detected optical afterglows, making the GRB-host 
associations robust.  Our results lend credence to the primary conclusions
of our previous study of short-duration GRB hosts and their redshifts, which
was based in part on XRT positions ($\leq 5\arcsec$), and in some cases
on probabilistic redshifts from host magnitudes.

We now attempt to estimate the contribution of high-redshift galaxies to the 
observed short-duration population.  We first consider only those short-duration
events with optical afterglows (and hence secure host identifications) and
secure host redshifts (from spectroscopy): GRBs 050709 ($z = 0.16$), 
050724 ($z = 0.26$), 051221A ($z = 0.55$), 061006 ($z = 0.44$; 
\citealt{bfp+07}), and the two hosts presented here.  High redshift ($z > 0.7$)
hosts constitute 33\% of this sample.  However, because these events are biased 
towards the brightest (and therefore, probabilistically, most nearby) hosts 
(see \citealt{bfp+07}, Fig.~8), we also consider three additional events
with secure hosts but no host redshifts (due primarily to faintness): GRBs
060313 \citep{rvp+06}, 060121 \citep{ltf+06,dcg+06}, and 051227 \citep{bfp+07}. 
If, as is likely from their faintness, all three events lie at $z > 0.7$, the 
percentage of high redshift short-duration GRBs may be as high as 56\%.  We 
note that this result does not change dramatically if we include XRT-host 
associations (e.g.~GRBs 050509B and 060502B).  The range of one- to two-thirds 
is in line with our previous analysis.  However, we caution that the errors
on these estimates are quite large due to small number statistics, and we 
do not consider here any selections effects associated with the above samples.  

At $z \sim 0.9$, the isotropic prompt energy release from GRB\,070429B 
is $E_{\gamma,\mathrm{iso}} =  1.4 \pm 0.2 \times 10^{50}$\,erg, while the
corresponding value for GRB\,070714B is $E_{\gamma,\mathrm{iso}} = 1.2 \pm 0.1 
\times 10^{51}$\,erg.  These values are several orders of magnitude larger than 
most of the previously established short-duration GRBs at $z < 0.5$, though 
still well short of typical values for long-duration events 
($E_{\gamma,\mathrm{iso}} \gtrsim 10^{52}$\,erg; \citealt{bkb+07}).  The X-ray 
luminosity at $t = 1$\,d, $L_{X,1}$, extrapolated from the online \Swift\ X-ray 
light curve repository \citep{ebp+07}, is consistent with the correlation found 
between $L_{X,1}$ and $E_{\gamma,\mathrm{iso}}$ found for the majority of 
short-duration events \citep{n07,b07}.  Like long-duration events, collimation 
corrections may play an important role in the total energy budget for 
short-duration bursts (e.g.~GRB\,051221A; \citealt{sbk+06,bgc+06}).  However, 
if a full 56\% of short-duration events reside at $z > 0.7$, the energy release 
of short-duration events may approach the canonical long-duration value of 
$E_{\gamma} \sim 10^{51}$\,erg \citep{fks+01,bkf03}.  Energies in excess of 
$10^{51}$\,erg are difficult to explain in the context of $\nu \bar{\nu}$ 
annihilation models \citep{rr02}, and may instead favor energy extraction via 
magneto-hydrodynamic processes (e.g.~\citealt{bz77}).

Much as with energetics, the specific star formation rate of GRB\,070429B
falls intermediate between the nearby short-duration events
($\phi \lesssim 1$\,M$_{\odot}$ yr$^{-1}$ L$_{*}^{-1}$; \citealt{n07} and
references therein) and long-duration GRBs ($\phi \sim 10$\,M$_{\odot}$ 
yr$^{-1}$ L$_{*}^{-1}$; \citealt{chg04}).  The specific star formation rate of
GRB\,070714B, however, is more consistent with the long-duration population, 
although large errors preclude any firm conclusions from being drawn.

The detection of short-duration GRBs in late-type galaxies at $z \gtrsim 1$ is 
more consistent with our current understanding of compact binary coalescence 
models (e.g.~\citealt{gp06,hgw+06,ngf06}).  Progenitor models with long delay 
times ($\tau \sim 6$\,Gyr) and narrow, log-normal distributions ($\sigma \sim
0.3$) were consistent with previous analyses based solely on the most nearby 
events.  Our results here favor models with a shorter characteristic delay
time ($\tau \sim 4$\,Gyr), but much broader distribution ($\sigma \sim 1$;
\citealt{ngf06}).  Further studies with a significantly expanded sample will 
be required for a detailed comparison of the properties of short-duration hosts 
with those predicted by compact binary merger models (e.g.~\citealt{bpb+06}).

\acknowledgments 
Based in part on observations obtained at the Gemini Observatory (Program ID:
GS-2007A-Q-6), which is operated by the Association of Universities for 
Research in Astronomy, Inc., under a cooperative agreement with the NSF on 
behalf of the Gemini partnership: the National Science Foundation (United 
States), the Science and Technology Facilities Council (United Kingdom), the 
National Research Council (Canada), CONICYT (Chile), the Australian Research 
Council (Australia), CNPq (Brazil) and SECYT (Argentina).  S.~B.~C.~is supported
by a NASA Graduate Student Research Fellowship.  M.~M.~K.~acknowledges funding
from the Gordon and Betty Moore Foundation.  GRB research at Caltech is
supported by NASA.

{\it Facilities:} \facility{Keck:I (LRIS)} \facility{Gemini:South (GMOS)}


\begin{thebibliography}{64}
\expandafter\ifx\csname natexlab\endcsname\relax\def\natexlab#1{#1}\fi

\bibitem[{{Antonelli} {et~al.}(2007){Antonelli}, {Stella}, {Tagliaferri},
  {Jehin}, \& {Schmidtobreick}}]{GCN.6372}
{Antonelli}, L.~A., {Stella}, L., {Tagliaferri}, G., {Jehin}, E., and
  {Schmidtobreick}, L. 2007, {GCN Circular} 6372

\bibitem[{{Barbier} {et~al.}(2007){Barbier}, {Barthelmy}, {Cummings},
  {Fenimore}, {Gehrels}, {Krimm}, {Markwardt}, {Palmer}, {Parsons}, {Racusin},
  {Sakamoto}, {Sato}, {Stamatikos}, {Tueller}, \& {Ukwatta}}]{GCN.6623}
{Barbier}, L. {\it et al.} 2007, {GCN Circular} 6623

\bibitem[{{Barthelmy} {et~al.}(2005{\natexlab{a}}){Barthelmy}, {Barbier},
  {Cummings}, {Fenimore}, {Gehrels}, {Hullinger}, {Krimm}, {Markwardt},
  {Palmer}, {Parsons}, {Sato}, {Suzuki}, {Takahashi}, {Tashiro}, \&
  {Tueller}}]{bbc+05}
{Barthelmy}, S.~D. {\it et al.} 2005{\natexlab{a}}, ArXiv Astrophysics e-prints

\bibitem[{{Barthelmy} {et~al.}(2005{\natexlab{b}}){Barthelmy}, {Chincarini},
  {Burrows}, {Gehrels}, {Covino}, {Moretti}, {Romano}, {O'Brien}, {Sarazin},
  {Kouveliotou}, {Goad}, {Vaughan}, {Tagliaferri}, {Zhang}, {Antonelli},
  {Campana}, {Cummings}, {D'Avanzo}, {Davies}, {Giommi}, {Grupe}, {Kaneko},
  {Kennea}, {King}, {Kobayashi}, {Melandri}, {Meszaros}, {Nousek}, {Patel},
  {Sakamoto}, \& {Wijers}}]{bcb+05}
---. 2005{\natexlab{b}}, \nat, 438, 994

\bibitem[{{Beardmore} {et~al.}(2007){Beardmore}, {Page}, {Evans}, {Starling},
  {Guidorzi}, \& {Markwardt}}]{GCN.6360}
{Beardmore}, A.~P., {Page}, K.~L., {Evans}, P.~A., {Starling}, R.~L.~C.,
  {Guidorzi}, C., and {Markwardt}, C. 2007, {GCN Circular} 6360

\bibitem[{{Beckwith} {et~al.}(2006){Beckwith}, {Stiavelli}, {Koekemoer},
  {Caldwell}, {Ferguson}, {Hook}, {Lucas}, {Bergeron}, {Corbin}, {Jogee},
  {Panagia}, {Robberto}, {Royle}, {Somerville}, \& {Sosey}}]{bsk+06}
{Beckwith}, S.~V.~W. {\it et al.} 2006, \aj, 132, 1729

\bibitem[{{Belczynski} {et~al.}(2006){Belczynski}, {Perna}, {Bulik},
  {Kalogera}, {Ivanova}, \& {Lamb}}]{bpb+06}
{Belczynski}, K., {Perna}, R., {Bulik}, T., {Kalogera}, V., {Ivanova}, N., and
  {Lamb}, D.~Q. 2006, \apj, 648, 1110

\bibitem[{{Berger}(2007)}]{b07}
{Berger}, E. 2007, \apj, 670, 1254

\bibitem[{{Berger} {et~al.}(2007){Berger}, {Fox}, {Price}, {Nakar}, {Gal-Yam},
  {Holz}, {Schmidt}, {Cucchiara}, {Cenko}, {Kulkarni}, {Soderberg}, {Frail},
  {Penprase}, {Rau}, {Ofek}, {Burnell}, {Cameron}, {Cowie}, {Dopita}, {Hook},
  {Peterson}, {Podsiadlowski}, {Roth}, {Rutledge}, {Sheppard}, \&
  {Songaila}}]{bfp+07}
{Berger}, E. {\it et al.} 2007, \apj, 664, 1000

\bibitem[{{Berger} {et~al.}(2005{\natexlab{a}}){Berger}, {Kulkarni}, {Fox},
  {Soderberg}, {Harrison}, {Nakar}, {Kelson}, {Gladders}, {Mulchaey}, {Oemler},
  {Dressler}, {Cenko}, {Price}, {Schmidt}, {Frail}, {Morrell}, {Gonzalez},
  {Krzeminski}, {Sari}, {Gal-Yam}, {Moon}, {Penprase}, {Jayawardhana},
  {Scholz}, {Rich}, {Peterson}, {Anderson}, {McNaught}, {Minezaki}, {Yoshii},
  {Cowie}, \& {Pimbblet}}]{bkf+05}
---. 2005{\natexlab{a}}, \apj, 634, 501

\bibitem[{{Berger} {et~al.}(2003){Berger}, {Kulkarni}, \& {Frail}}]{bkf03}
{Berger}, E., {Kulkarni}, S.~R., and {Frail}, D.~A. 2003, \apj, 590, 379

\bibitem[{{Berger} {et~al.}(2005{\natexlab{b}}){Berger}, {Price}, {Cenko},
  {Gal-Yam}, {Soderberg}, {Kasliwal}, {Leonard}, {Cameron}, {Frail},
  {Kulkarni}, {Murphy}, {Krzeminski}, {Piran}, {Lee}, {Roth}, {Moon}, {Fox},
  {Harrison}, {Persson}, {Schmidt}, {Penprase}, {Rich}, {Peterson}, \&
  {Cowie}}]{bpc+05}
{Berger}, E. {\it et al.} 2005{\natexlab{b}}, \nat, 438, 988

\bibitem[{{Bessell}(1999)}]{b99}
{Bessell}, M.~S. 1999, \pasp, 111, 1426

\bibitem[{{Blandford} \& {Znajek}(1977)}]{bz77}
{Blandford}, R.~D. and {Znajek}, R.~L. 1977, \mnras, 179, 433

\bibitem[{{Bloom} {et~al.}(2006){Bloom}, {Prochaska}, {Pooley}, {Blake},
  {Foley}, {Jha}, {Ramirez-Ruiz}, {Granot}, {Filippenko}, {Sigurdsson},
  {Barth}, {Chen}, {Cooper}, {Falco}, {Gal}, {Gerke}, {Gladders}, {Greene},
  {Hennanwi}, {Ho}, {Hurley}, {Koester}, {Li}, {Lubin}, {Newman}, {Perley},
  {Squires}, \& {Wood-Vasey}}]{bpp+06}
{Bloom}, J.~S. {\it et al.} 2006, \apj, 638, 354

\bibitem[{{Bohlin} \& {Gilliland}(2004)}]{bg04}
{Bohlin}, R.~C. and {Gilliland}, R.~L. 2004, \aj, 128, 3053

\bibitem[{{Burrows} {et~al.}(2006){Burrows}, {Grupe}, {Capalbi}, {Panaitescu},
  {Patel}, {Kouveliotou}, {Zhang}, {M{\'e}sz{\'a}ros}, {Chincarini}, {Gehrels},
  \& {Wijers}}]{bgc+06}
{Burrows}, D.~N. {\it et al.} 2006, \apj, 653, 468

\bibitem[{{Burrows} {et~al.}(2005){Burrows}, {Hill}, {Nousek}, {Kennea},
  {Wells}, {Osborne}, {Abbey}, {Beardmore}, {Mukerjee}, {Short}, {Chincarini},
  {Campana}, {Citterio}, {Moretti}, {Pagani}, {Tagliaferri}, {Giommi},
  {Capalbi}, {Tamburelli}, {Angelini}, {Cusumano}, {Braeuninger}, {Burkert}, \&
  {Hartner}}]{bhn+05}
---. 2005, ArXiv Astrophysics e-prints

\bibitem[{{Butler} {et~al.}(2007){Butler}, {Kocevski}, {Bloom}, \&
  {Curtis}}]{bkb+07}
{Butler}, N.~R., {Kocevski}, D., {Bloom}, J.~S., and {Curtis}, J.~L. 2007,
  \apj, 671, 656

\bibitem[{{Christensen} {et~al.}(2004){Christensen}, {Hjorth}, \&
  {Gorosabel}}]{chg04}
{Christensen}, L., {Hjorth}, J., and {Gorosabel}, J. 2004, \aap, 425, 913

\bibitem[{{Covino} {et~al.}(2007){Covino}, {Piranomonte}, {D'Avanzo},
  {Antonelli}, {Luise}, {Pedani}, \& {Alfonso}}]{GCN.6635}
{Covino}, S., {Piranomonte}, S., {D'Avanzo}, P., {Antonelli}, L.~A., {Luise},
  F.~d., {Pedani}, M., and {Alfonso}, N.~P. 2007, {GCN Circular} 6635

\bibitem[{{Cucchiara} {et~al.}(2007){Cucchiara}, {Cenko}, {Fox}, {Berger}, \&
  {Bloom}}]{GCN.6370}
{Cucchiara}, A., {Cenko}, S.~B., {Fox}, D.~B., {Berger}, E., and {Bloom}, J.~S.
  2007, {GCN Circular} 6370

\bibitem[{{de Ugarte Postigo} {et~al.}(2006){de Ugarte Postigo},
  {Castro-Tirado}, {Guziy}, {Gorosabel}, {J{\'o}hannesson}, {Aloy}, {McBreen},
  {Lamb}, {Benitez}, {Jel{\'{\i}}nek}, {Pandey}, {Coe},
  {P{\'e}rez-Ram{\'{\i}}rez}, {Aceituno}, {Alises}, {Acosta-Pulido},
  {G{\'o}mez}, {L{\'o}pez}, {Donaghy}, {Nakagawa}, {Sakamoto}, {Ricker},
  {Hearty}, {Bayliss}, {Gyuk}, \& {York}}]{dcg+06}
{de Ugarte Postigo}, A. {\it et al.} 2006, \apjl, 648, L83

\bibitem[{{Eichler} {et~al.}(1989){Eichler}, {Livio}, {Piran}, \&
  {Schramm}}]{elp+89}
{Eichler}, D., {Livio}, M., {Piran}, T., and {Schramm}, D.~N. 1989, \nat, 340,
  126

\bibitem[{{Evans} {et~al.}(2007){Evans}, {Beardmore}, {Page}, {Tyler},
  {Osborne}, {Goad}, {O'Brien}, {Vetere}, {Racusin}, {Morris}, {Burrows},
  {Capalbi}, {Perri}, {Gehrels}, \& {Romano}}]{ebp+07}
{Evans}, P.~A. {\it et al.} 2007, \aap, 469, 379

\bibitem[{{Fox} {et~al.}(2005){Fox}, {Frail}, {Price}, {Kulkarni}, {Berger},
  {Piran}, {Soderberg}, {Cenko}, {Cameron}, {Gal-Yam}, {Kasliwal}, {Moon},
  {Harrison}, {Nakar}, {Schmidt}, {Penprase}, {Chevalier}, {Kumar}, {Roth},
  {Watson}, {Lee}, {Shectman}, {Phillips}, {Roth}, {McCarthy}, {Rauch},
  {Cowie}, {Peterson}, {Rich}, {Kawai}, {Aoki}, {Kosugi}, {Totani}, {Park},
  {MacFadyen}, \& {Hurley}}]{ffp+05}
{Fox}, D.~B. {\it et al.} 2005, \nat, 437, 845

\bibitem[{{Frail} {et~al.}(2001){Frail}, {Kulkarni}, {Sari}, {Djorgovski},
  {Bloom}, {Galama}, {Reichart}, {Berger}, {Harrison}, {Price}, {Yost},
  {Diercks}, {Goodrich}, \& {Chaffee}}]{fks+01}
{Frail}, D.~A. {\it et al.} 2001, \apjl, 562, L55

\bibitem[{{Gehrels} {et~al.}(2006){Gehrels}, {Norris}, {Barthelmy}, {Granot},
  {Kaneko}, {Kouveliotou}, {Markwardt}, {M{\'e}sz{\'a}ros}, {Nakar}, {Nousek},
  {O'Brien}, {Page}, {Palmer}, {Parsons}, {Roming}, {Sakamoto}, {Sarazin},
  {Schady}, {Stamatikos}, \& {Woosley}}]{gnb+06}
{Gehrels}, N. {\it et al.} 2006, \nat, 444, 1044

\bibitem[{{Gehrels} {et~al.}(2005){Gehrels}, {Sarazin}, {O'Brien}, {Zhang},
  {Barbier}, {Barthelmy}, {Blustin}, {Burrows}, {Cannizzo}, {Cummings}, {Goad},
  {Holland}, {Hurkett}, {Kennea}, {Levan}, {Markwardt}, {Mason}, {Meszaros},
  {Page}, {Palmer}, {Rol}, {Sakamoto}, {Willingale}, {Angelini}, {Beardmore},
  {Boyd}, {Breeveld}, {Campana}, {Chester}, {Chincarini}, {Cominsky},
  {Cusumano}, {de Pasquale}, {Fenimore}, {Giommi}, {Gronwall}, {Grupe}, {Hill},
  {Hinshaw}, {Hjorth}, {Hullinger}, {Hurley}, {Klose}, {Kobayashi},
  {Kouveliotou}, {Krimm}, {Mangano}, {Marshall}, {McGowan}, {Moretti},
  {Mushotzky}, {Nakazawa}, {Norris}, {Nousek}, {Osborne}, {Page}, {Parsons},
  {Patel}, {Perri}, {Poole}, {Romano}, {Roming}, {Rosen}, {Sato}, {Schady},
  {Smale}, {Sollerman}, {Starling}, {Still}, {Suzuki}, {Tagliaferri},
  {Takahashi}, {Tashiro}, {Tueller}, {Wells}, {White}, \& {Wijers}}]{gso+05}
---. 2005, \nat, 437, 851

\bibitem[{{Graham} {et~al.}(2007){Graham}, {Fruchter}, {Levan}, {Nysewander},
  {Tanvir}, {Dahlen}, {Bersier}, \& {Pe'er}}]{GCN.6836}
{Graham}, J.~F., {Fruchter}, A.~S., {Levan}, A.~J., {Nysewander}, M., {Tanvir},
  N.~R., {Dahlen}, T., {Bersier}, D., and {Pe'er}, A. 2007, {GCN Circular} 6836

\bibitem[{{Grupe} {et~al.}(2006){Grupe}, {Burrows}, {Patel}, {Kouveliotou},
  {Zhang}, {M{\'e}sz{\'a}ros}, {Wijers}, \& {Gehrels}}]{gbp+06}
{Grupe}, D., {Burrows}, D.~N., {Patel}, S.~K., {Kouveliotou}, C., {Zhang}, B.,
  {M{\'e}sz{\'a}ros}, P., {Wijers}, R.~A.~M., and {Gehrels}, N. 2006, \apj,
  653, 462

\bibitem[{{Guetta} \& {Piran}(2006)}]{gp06}
{Guetta}, D. and {Piran}, T. 2006, \aap, 453, 823

\bibitem[{{Hjorth} {et~al.}(2005){Hjorth}, {Watson}, {Fynbo}, {Price},
  {Jensen}, {J{\o}rgensen}, {Kubas}, {Gorosabel}, {Jakobsson}, {Sollerman},
  {Pedersen}, \& {Kouveliotou}}]{hwf+05}
{Hjorth}, J. {\it et al.} 2005, \nat, 437, 859

\bibitem[{{Holland} {et~al.}(2007){Holland}, {Pasquale}, \&
  {Markwardt}}]{GCN.7145}
{Holland}, S.~T., {Pasquale}, M.~d., and {Markwardt}, C.~. 2007, {GCN Circular}
  7145

\bibitem[{{Hook} {et~al.}(2003){Hook}, {Allington-Smith}, {Beard}, {Crampton},
  {Davies}, {Dickson}, {Ebbers}, {Fletcher}, {Jorgensen}, {Jean}, {Juneau},
  {Murowinski}, {Nolan}, {Laidlaw}, {Leckie}, {Marshall}, {Purkins},
  {Richardson}, {Roberts}, {Simons}, {Smith}, {Stilburn}, {Szeto}, {Tierney},
  {Wolff}, \& {Wooff}}]{hab+03}
{Hook}, I. {\it et al.} 2003, in Instrument Design and Performance for
  Optical/Infrared Ground-based Telescopes. Edited by Iye, Masanori; Moorwood,
  Alan F. M. Proceedings of the SPIE, Volume 4841, pp. 1645-1656 (2003)., ed.
  M.~{Iye} \& A.~F.~M. {Moorwood}, 1645--1656

\bibitem[{{Hopman} {et~al.}(2006){Hopman}, {Guetta}, {Waxman}, \& {Portegies
  Zwart}}]{hgw+06}
{Hopman}, C., {Guetta}, D., {Waxman}, E., and {Portegies Zwart}, S. 2006,
  \apjl, 643, L91

\bibitem[{{Horne}(1986)}]{h86}
{Horne}, K. 1986, \pasp, 98, 609

\bibitem[{{Jakobsson} {et~al.}(2006){Jakobsson}, {Levan}, {Fynbo}, {Priddey},
  {Hjorth}, {Tanvir}, {Watson}, {Jensen}, {Sollerman}, {Natarajan},
  {Gorosabel}, {Castro Cer{\'o}n}, {Pedersen}, {Pursimo}, {{\'A}rnad{\'o}ttir},
  {Castro-Tirado}, {Davis}, {Deeg}, {Fiuza}, {Mykolaitis}, \& {Sousa}}]{jlf+06}
{Jakobsson}, P. {\it et al.} 2006, \aap, 447, 897

\bibitem[{{Kennicutt}(1998)}]{k98}
{Kennicutt}, Jr., R.~C. 1998, \araa, 36, 189

\bibitem[{{Levan} {et~al.}(2007){Levan}, {Tanvir}, {Bonfield},
  {Martinez-Sansigre}, {Graham}, \& {Fruchter}}]{GCN.6630}
{Levan}, A.~J., {Tanvir}, N.~R., {Bonfield}, D., {Martinez-Sansigre}, A.,
  {Graham}, J., and {Fruchter}, A. 2007, {GCN Circular} 6630

\bibitem[{{Levan} {et~al.}(2006){Levan}, {Tanvir}, {Fruchter}, {Rol}, {Fynbo},
  {Hjorth}, {Williams}, {Bergeron}, {Bersier}, {Bremer}, {Grav}, {Jakobsson},
  {Nilsson}, {Olszewski}, {Priddey}, {Rafferty}, \& {Rhoads}}]{ltf+06}
{Levan}, A.~J. {\it et al.} 2006, \apjl, 648, L9

\bibitem[{{Markwardt} {et~al.}(2007){Markwardt}, {Barbier}, {Barthelmy},
  {Beardmore}, {Cannizzo}, {Chester}, {Gehrels}, {Hunsberger}, {Landsman},
  {Page}, {Palmer}, {Sato}, \& {Starling}}]{GCN.6358}
{Markwardt}, C.~B. {\it et al.} 2007, {GCN Circular} 6358

\bibitem[{{Melandri}(2007)}]{GCN.6621}
{Melandri}, A. 2007, {GCN Circular} 6621

\bibitem[{{Nakar}(2007)}]{n07}
{Nakar}, E. 2007, \physrep, 442, 166

\bibitem[{{Nakar} {et~al.}(2006){Nakar}, {Gal-Yam}, \& {Fox}}]{ngf06}
{Nakar}, E., {Gal-Yam}, A., and {Fox}, D.~B. 2006, \apj, 650, 281

\bibitem[{{Norris} {et~al.}(2007){Norris}, {Barthelmy}, \&
  {Ukwatta}}]{GCN.6631}
{Norris}, J., {Barthelmy}, S.~D., and {Ukwatta}, T. 2007, {GCN Circular} 6631

\bibitem[{{Norris} \& {Bonnell}(2006)}]{nb06}
{Norris}, J.~P. and {Bonnell}, J.~T. 2006, \apj, 643, 266

\bibitem[{{Nysewander} {et~al.}(2007){Nysewander}, {Fruchter}, \&
  {Graham}}]{GCN.6604}
{Nysewander}, M., {Fruchter}, A., and {Graham}, J. 2007, {GCN Circular} 6604

\bibitem[{{Oke} {et~al.}(1995){Oke}, {Cohen}, {Carr}, {Cromer}, {Dingizian},
  {Harris}, {Labrecque}, {Lucinio}, {Schaal}, {Epps}, \& {Miller}}]{occ+95}
{Oke}, J.~B. {\it et al.} 1995, \pasp, 107, 375

\bibitem[{{Oke} \& {Gunn}(1983)}]{og83}
{Oke}, J.~B. and {Gunn}, J.~E. 1983, \apj, 266, 713

\bibitem[{{Perley} {et~al.}(2007{\natexlab{a}}){Perley}, {Bloom}, {Modjaz},
  {Poznanski}, \& {Thoene}}]{GCN.7140}
{Perley}, D.~A., {Bloom}, J.~S., {Modjaz}, M., {Poznanski}, D., and {Thoene},
  C.~C. 2007{\natexlab{a}}, {GCN Circular} 7140

\bibitem[{{Perley} {et~al.}(2007{\natexlab{b}}){Perley}, {Bloom}, {Thoene}, \&
  {Butler}}]{GCN.6652}
{Perley}, D.~A., {Bloom}, J.~S., {Thoene}, C., and {Butler}, N.~R.
  2007{\natexlab{b}}, {GCN Circular} 6652

\bibitem[{{Racusin} {et~al.}(2007{\natexlab{a}}){Racusin}, {Kennea}, {Pagani},
  {Vetere}, \& {Evans}}]{GCN.6627}
{Racusin}, J., {Kennea}, J., {Pagani}, C., {Vetere}, L., and {Evans}, P.
  2007{\natexlab{a}}, {GCN Circular} 6627

\bibitem[{{Racusin} {et~al.}(2007{\natexlab{b}}){Racusin}, {Barthelmy},
  {Burrows}, {Chester}, {Gehrels}, {Krimm}, {Palmer}, \& {Sakamoto}}]{GCN.6620}
{Racusin}, J.~L., {Barthelmy}, S.~D., {Burrows}, D.~N., {Chester}, M.~M.,
  {Gehrels}, N., {Krimm}, H.~A., {Palmer}, D.~M., and {Sakamoto}, T.
  2007{\natexlab{b}}, {GCN Circular} 6620

\bibitem[{{Roming} {et~al.}(2005){Roming}, {Kennedy}, {Mason}, {Nousek}, {Ahr},
  {Bingham}, {Broos}, {Carter}, {Hancock}, {Huckle}, {Hunsberger}, {Kawakami},
  {Killough}, {Koch}, {McLelland}, {Smith}, {Smith}, {Soto}, {Boyd},
  {Breeveld}, {Holland}, {Ivanushkina}, {Pryzby}, {Still}, \& {Stock}}]{rkm+05}
{Roming}, P.~W.~A. {\it et al.} 2005, ArXiv Astrophysics e-prints

\bibitem[{{Roming} {et~al.}(2006){Roming}, {Vanden Berk}, {Pal'shin}, {Pagani},
  {Norris}, {Kumar}, {Krimm}, {Holland}, {Gronwall}, {Blustin}, {Zhang},
  {Schady}, {Sakamoto}, {Osborne}, {Nousek}, {Marshall}, {M{\'e}sz{\'a}ros},
  {Golenetskii}, {Gehrels}, {Frederiks}, {Campana}, {Burrows}, {Boyd},
  {Barthelmy}, \& {Aptekar}}]{rvp+06}
---. 2006, \apj, 651, 985

\bibitem[{{Rosswog} \& {Ramirez-Ruiz}(2002)}]{rr02}
{Rosswog}, S. and {Ramirez-Ruiz}, E. 2002, \mnras, 336, L7

\bibitem[{{Schlegel} {et~al.}(1998){Schlegel}, {Finkbeiner}, \&
  {Davis}}]{sfd98}
{Schlegel}, D.~J., {Finkbeiner}, D.~P., and {Davis}, M. 1998, \apj, 500, 525

\bibitem[{{Soderberg} {et~al.}(2006){Soderberg}, {Berger}, {Kasliwal}, {Frail},
  {Price}, {Schmidt}, {Kulkarni}, {Fox}, {Cenko}, {Gal-Yam}, {Nakar}, \&
  {Roth}}]{sbk+06}
{Soderberg}, A.~M. {\it et al.} 2006, \apj, 650, 261

\bibitem[{{Spergel} {et~al.}(2007){Spergel}, {Bean}, {Dor{\'e}}, {Nolta},
  {Bennett}, {Dunkley}, {Hinshaw}, {Jarosik}, {Komatsu}, {Page}, {Peiris},
  {Verde}, {Halpern}, {Hill}, {Kogut}, {Limon}, {Meyer}, {Odegard}, {Tucker},
  {Weiland}, {Wollack}, \& {Wright}}]{sbd+07}
{Spergel}, D.~N. {\it et al.} 2007, \apjs, 170, 377

\bibitem[{{Stone}(1996)}]{s96}
{Stone}, R.~P.~S. 1996, \apjs, 107, 423

\bibitem[{{Tueller} {et~al.}(2007){Tueller}, {Barbier}, {Barthelmy},
  {Cummings}, {Fenimore}, {Gehrels}, {Krimm}, {Markwardt}, {Palmer}, {Parsons},
  {Sakamoto}, {Sato}, \& {Stamatikos}}]{GCN.6365}
{Tueller}, J. {\it et al.} 2007, {GCN Circular} 6365

\bibitem[{{van Dokkum}(2001)}]{v01}
{van Dokkum}, P.~G. 2001, \pasp, 113, 1420

\bibitem[{{Villasenor} {et~al.}(2005){Villasenor}, {Lamb}, {Ricker}, {Atteia},
  {Kawai}, {Butler}, {Nakagawa}, {Jernigan}, {Boer}, {Crew}, {Donaghy}, {Doty},
  {Fenimore}, {Galassi}, {Graziani}, {Hurley}, {Levine}, {Martel}, {Matsuoka},
  {Olive}, {Prigozhin}, {Sakamoto}, {Shirasaki}, {Suzuki}, {Tamagawa},
  {Vanderspek}, {Woosley}, {Yoshida}, {Braga}, {Manchanda}, {Pizzichini},
  {Takagishi}, \& {Yamauchi}}]{vlr+05}
{Villasenor}, J.~S. {\it et al.} 2005, \nat, 437, 855

\end{thebibliography}

\end{document}